\documentclass[fleqn,twoside]{article}
\usepackage{espcrc2}

\usepackage{graphicx}
\usepackage[figuresright]{rotating}

\newcommand{\AmS}{{\protect\the\textfont2
  A\kern-.1667em\lower.5ex\hbox{M}\kern-.125emS}}
\newcommand{\<}{\langle}
\renewcommand{\>}{\rangle}
\title{Lattice QCD at maximal twist}

\author{R. Frezzotti\address[URTV]
{Dipartimento di Fisica, Universit\`a di Roma 
{\it Tor Vergata} and INFN Sezione di {\it Tor Vergata}\\
Via della Ricerca Scientifica - 00133 Roma (ITALY)} and 
        G.C. Rossi\addressmark[URTV]
        \thanks{Speaker}\thanks{G.C.R. wishes to thank NIC at Desy-Zeuthen
                and the Humboldt Foundation for financial support.}
                }
       
\begin{document}

\begin{abstract}
In this review we discuss the general features of maximally twisted lattice QCD. 
In particular, we illustrate how automatic O($a$) improvement can be achieved and 
how it is possible to set up a lattice regularization scheme where the problem 
of wrong chirality mixing (be it finite or infinite) affecting the computation 
of the matrix elements of the ${\cal{CP}}$-conserving effective weak Hamiltonian 
is neatly avoided, while having at the same time a positive determinant even for 
non-degenerate quark pairs. The question of reducing the large cutoff effects that 
appear when the quark mass tends to zero as a consequence of parity and iso-spin 
breaking in the action is also addressed. It is shown that such dangerous lattice 
artifacts are strongly suppressed if the clover term is added to the action or, 
alternatively, the critical mass is chosen so as to enforce the restoration of 
parity. 

\vspace{1pc}
\end{abstract}

\maketitle

\section{Introduction and content}
\label{sec:INTRO}

Waiting for new generation computers that might allow full-fledged simulations 
with exactly chirally invariant fermions, i.e.\ fermions obeying the Ginsparg--Wilson
condition~\cite{GW}, a viable alternative could be to use maximally twisted Wilson 
fermions~\cite{TM,FR1,FRC,FR2,FMPR}, possibly coupled with a judicious choice 
of the gauge action~\cite{MONT}. 

In this review we wish to outline the structure and the properties 
of maximally twisted lattice QCD (Mtm-LQCD) as developed in refs.~\cite{FR1,FRC,FR2,FMPR}.
Soon after introducing the idea that to avoid exceptional configurations
in Wilson fermion simulations one should introduce quarks in flavour pairs 
and have the Wilson term rotated with respect to the quark mass term 
by an axial rotation in iso-spin space, it was realized that an 
especially useful choice for that angle is to set it at its maximal
value, $|\omega|=\pi/2$, because in this situation O($a$) (actually 
O($a^{2k+1}$), $k\geq 0$) improvement of 
physical quantities is automatic (sect.~\ref{sec:IMPR}). 

It was then shown in~\cite{FRC} that the nice improvement properties enjoyed by 
Mtm-LQCD, derived for pairs of mass degenerate quarks in~\cite{FR1}, can be 
immediately extended to the more interesting non-degenerate case, without 
loosing the positivity of the corresponding fermion determinant 
(sect.~\ref{sec:NONDEG}). This last property is obviously crucial if one 
wants to be able to set up workable Monte Carlo-like simulation algorithms for QCD.

With these ingredients and exploiting the flexibility offered by the possibility 
of regularizing different valence flavours with different value of the Wilson 
parameter, $r$, it is possible to construct~\cite{FRC} a hybrid theory, where sea quarks 
are introduced as pairs of non-degenerate particles and valence quarks are regularized 
as \"Osterwalder--Seiler~\cite{OS} fermions, such that no ``wrong chirality'' 
mixing~\cite{BMMRT} affects the computation of the matrix elements of the 
${\cal{CP}}$-conserving $\Delta S=1,2$ effective weak Hamiltonian 
(sect.~\ref{sec:NOMIX}). 

Despite all these nice features, there remain important cutoff effects, 
originating from the breaking of parity and iso-spin induced by the presence 
of the twisted Wilson term in the action, which tend to become large as the 
quark mass is lowered. These lattice artifacts have been discussed both 
in chiral perturbation theory ($\chi$PT)~\cite{SHWUNEW}-\cite{SH05}, 
as well as in the language of the Symanzik expansion~\cite{FMPR} with the 
conclusion that they can be substantially reduced 
if the clover term~\cite{SW} is introduced in the action with its 
non-perturbatively fixed $c_{SW}$ coefficient~\cite{LU} or, alternatively, if 
the critical mass is chosen in some ``optimal way'' (sect.~\ref{sec:OPTCH}).

For lack of space we will not discuss how and to what extent the strategy 
outlined above for improvement can be extended to the Schr\"odinger functional 
formalism~\cite{FR3,SSF}. Nor we will address the very important issue of 
meta-stabilities~\cite{META1,META2,CPT}, that are seen to affect unquenched 
simulations~\cite{FARC} at coarse lattice spacing, and the closely 
related flavour breaking effects visible in the 
value of the square mass difference between charged and neutral 
pions~\cite{CHRMI}. These questions and the the present status of quenched and 
unquenched twisted simulations have been recently reviewed in ref.~\cite{SHI}. 
In the concluding section (sect.~\ref{sec:CONCL}) we will only offer some hints 
on possible ways to overcome (even on coarse lattice spacings) difficulties 
associated with the existence of metastable phases and flavour breaking effects.

\section{Improvement}
\label{sec:IMPR}

The twisted fermion action for a pair of degenerate quarks in the 
``physical basis'' (where the parameter which gives a non-vanishing 
mass to pions, $m_q$, is real) has the expression 
\begin{equation}
S_{\rm{F}}^{\omega}\!=\!a^4 \!\sum_x\bar\psi(x)
[\gamma\tilde\nabla \!+\! W_{\rm{cr}}e^{-i\omega\gamma_5\tau_3}
\!+m_q]\psi(x)\, .\label{STM}
\end{equation}
In eq.~(\ref{STM}) we have introduced the following definitions: 
$\gamma\tilde\nabla=\frac{1}{2}\gamma_\mu(\nabla^\star_\mu+\nabla_\mu)$ 
with $\nabla_\mu$ and $\nabla^\star_\mu$ the forward and backward 
lattice covariant derivatives, respectively, and 
$W_{\rm{cr}}=-r\frac{a}{2}\nabla^\star_\mu\nabla_\mu+M_{\rm{cr}}(r)$
with $M_{\rm{cr}}(r)$ the critical mass. We observe 
that to match the $r$-parity property of the Wilson term $M_{\rm{cr}}(r)$ is 
necessarily an odd function of $r$. The form of the fermionic action in the 
``twisted basis'' is obtained by performing on the fermion fields the axial 
rotation 
\begin{equation}
\psi\rightarrow e^{i{\omega}\gamma_5\tau_3/2}\psi\, ,\quad
\bar\psi\rightarrow \bar\psi e^{i{\omega}\gamma_5\tau_3/2}\, .\label{ROT}
\end{equation}
The first observation about $S_{\rm{F}}^{\omega}$ (and the one that originally led 
to its introduction) is that the determinant of the associated Dirac--Wilson 
operator is positive definite for any $\omega\neq 0$ and $m_q\neq 0$ as 
its expression is seen to be given by 
\begin{eqnarray}
\hspace{-.5cm}&&{\cal{D}}^\omega_F={\rm{det}}[(D_{\rm{W}}^{\rm{cr}}+
m_q\cos\omega)^\dagger (D_{\rm{W}}^{\rm{cr}}+m_q\cos\omega)+
\nonumber\\
\hspace{-.5cm}&&+{m_q^2\sin^2\omega}]\, ,\quad 
D_{\rm{W}}^{\rm{cr}}= \gamma\tilde\nabla+W_{\rm{cr}}\, .\label{DET}
\end{eqnarray}
The lattice action~(\ref{STM}) is invariant under the (spurionic) 
transformations

\noindent 1) ${\cal{R}}_5 \times (r\to -r)\times (m_q\to -m_q)$, where 
\begin{eqnarray}
{{{\cal{R}}_5\!:\! \left \{\begin{array}{lll}
\hspace{-.1cm}\psi &\rightarrow & \psi'=\gamma_5 \psi\\
\hspace{-.1cm}\bar{\psi} &\rightarrow & \bar{\psi}'=-\bar{\psi}\gamma_5
\end{array}\right .} \quad
\qquad {\cal{R}}_5^2 = 1 } \label{R5DEF} \end{eqnarray}
2) ${\cal D}_d\times (r\to -r)\times (m_q\to -m_q)$, with
\begin{eqnarray}
{{\cal{D}}_d\!:\!\left \{\begin{array}{lll}
\hspace{-.1cm}U_\mu(x)\!&\!\!\!\!\rightarrow \!\!\!&\!U_\mu^\dagger(-x-a\hat\mu))
\label{DDDEF} \\
\hspace{-.1cm}(\psi(x),\bar{\psi}(x))\!&\!\!\!\!\rightarrow\!\!\!& \!e^{3i\pi/2}\, 
(\psi(-x),\bar{\psi}(-x))
\end{array}\right .} \end{eqnarray}
These invariances imply for the lattice vacuum expectation values (v.e.v.'s) 
of (multi-local) operators, $O(x)=O(x_1,x_2,\ldots,x_n)$, 
$x_1\neq x_2\neq\ldots\neq x_n$, the relations 
\begin{eqnarray}
\hspace{-.7cm}&&1)\, {{\langle O(x)\rangle\Big{|}^{({\omega})}_{({r,m_q})} 
= (-1)^{{P_{{\cal{R}}_5}[O]}}\langle
O(x)\rangle\Big{|}^{({\omega})}_{({-r,-m_q})}}} \label{R5INV}\\
\hspace{-.7cm}&&2)\, \langle  O(x)
\rangle \Big{|}^{({\omega})}_{({r,m_q})}=(-1)^{d[O]} \langle  O(\!{-}x)
\rangle \Big{|}^{({\omega})}_{({-r,-m_q})}\, , \label{DDINV}\end{eqnarray}
where $P_{{{\cal{R}}_5}[O]}$ is the ${{\cal{R}}_5}$-parity of $O$ and $d[O]$ 
its mass dimension.

We now prove that i) for generic values of the twisting angle, $\omega$, 
averages (Wilson average - WA) of v.e.v.'s computed with opposite 
values of $r$ are free of O($a^{2k+1}$), $k\geq 0$, cutoff effects; 
ii) for the special value $|\omega|=\pi/2$ averaging is unnecessary.

These results follow from using the Symanzik idea~\cite{SYMA} that lattice
artifacts can be described in terms of the continuum correlators 
of an effective low energy theory renormalized at the scale 
$a^{-1}$, as well as the symmetry properties of the 
lattice theory and the corresponding ones enjoyed by the associated 
continuum theory.

i) $|\omega|\neq\pi/2$ - We write the Symanzik expansion of the lattice 
v.e.v.\ of a multi-local, multiplicatively renormalizable (m.r.) operator
in the schematic form
\begin{eqnarray}
\hspace{-.7cm}&&\langle O(x)\rangle\Big{|}^{(\omega)}_{(r,m_q)}= 
\Big{[}\zeta^{O}_{O}(\omega,r)\langle O(x)\rangle +\nonumber\\
\hspace{-.7cm}&&+ a\sum_{\ell}\eta^{O}_{O_{\ell}}(\omega,r)
\langle O_{\ell}(x)\rangle\Big{]}^{\rm{cont}}_{(m_q)}+{\rm{O}}(a^2)
\, , \label{SYMEXP1}\end{eqnarray}
where the operators $O_{\ell}$ have mass dimension of one unit larger 
than that of $O$, $d[O_{\ell}]\!=\!d[O]\!+\!1$. We explicitly remark that factors 
of $m_q$ may appear in $O_{\ell}$. The lattice invariance~(\ref{DDINV}) 
allow us to write 
\begin{eqnarray}
\hspace{-.7cm}&&\langle O(x)\rangle\Big{|}^{(\omega)}_{(r,m_q)}= 
(-1)^{d[O]}\langle O(-x)\rangle\Big{|}^{(\omega)}_{(-r,-m_q)}=\nonumber\\
\hspace{-.7cm}&&=(-1)^{d[O]}\Big{[}\zeta^{O}_{O}(\omega,-r)\langle O(-x)\rangle+\nonumber\\
\hspace{-.7cm}&&+ a\sum_{\ell}\eta^{O}_{O_{\ell}}(\omega,-r)
\langle O_{\ell}(-x)\rangle\Big{]}^{\rm{cont}}_{(-m_q)}+{\rm{O}}(a^2)
\, . \label{SYMEXP2}\end{eqnarray}
We can now use the invariance ${\cal D}_d\times (m_q\to -m_q)$ of the 
continuum theory to bring continuum correlators computed in $-x$ and 
$-m_q$ to correlators computed in $x$ and $m_q$. One thus gets 
\begin{eqnarray}
\hspace{-.7cm}&&\langle O(x)\rangle\Big{|}^{(\omega)}_{(r,m_q)}
=\Big{[}\zeta^{O}_{O}(\omega,-r)\langle O(x) \rangle+
\nonumber\\
\hspace{-.7cm}&&-a\sum_{\ell}\eta^{O}_{O_{\ell}}(\omega,-r)
\langle O_{\ell}(x)\rangle\Big{]}^{\rm{cont}}_{(m_q)}\!\!+{\rm{O}}(a^2)
\, . \!\!\label{SYMEXP3}\end{eqnarray}
Comparison with~(\ref{SYMEXP1}) gives the relations 
\begin{eqnarray}
\zeta^{O}_{O}(\omega,r)=\zeta^{O}_{O}(\omega,-r)\, ,\,\,
\eta^{O}_{O_{\ell}}(\omega,r)=-\eta^{O}_{O_{\ell}}(\omega,-r)\, ,
\nonumber\end{eqnarray}
which imply the WA improvement formula 
\begin{eqnarray}
&&\langle O(x)\rangle \Big{|}^{({\omega})}_{({r},m_q)}+\langle  
O(x) \rangle \Big{|}^{({\omega})}_{({-r},m_q)}=\nonumber\\
&&=2\zeta^{O}_{O}(\omega,r)\langle O(x)\rangle\Big{|}^{\rm cont}_{(m_q)}
+{ {\rm{O}}(a^2)}\, .
\label{WA}\end{eqnarray}
Exploiting the ${\cal{R}}_5$-parity transformation properties of $O$, 
the previous formula can be rewritten in the form of a mass average (MA)
\begin{eqnarray}
&&\langle O(x)\rangle \Big{|}^{({\omega})}_{({r},m_q)}+(-1)^{{P_{{\cal{R}}_5}[O]}}
\langle O(x)\rangle \Big{|}^{({\omega})}_{({r},-m_q)}=\nonumber\\
&&=2\zeta^{O}_{O}(\omega,r)\langle O(x)\rangle\Big{|}^{\rm cont}_{(m_q)}
+{{\rm{O}}(a^2)}\, .
\label{MA}\end{eqnarray}
Note that in the presence of spontaneous chiral symmetry 
breaking the two terms in the l.h.s.\ of~(\ref{MA}) will be in general 
different even in the limit $m_q\to 0$~\footnote{In the absence of spontaneous 
chiral symmetry breaking, like in 2-dimensions, lattice data show 
improvement at any $\omega$ even without WA~\cite{DMJ}.}.

One can easily convince oneself that the  whole argument developed above 
is actually valid for any odd powers of $a$ in the Symanzik expansion. 
As a result all O($a^{2k+1}$), $k\geq 0$, terms get cancel by taking either the
Wilson or the mass average of lattice correlators.

ii) $|\omega|=\pi/2$ - At maximal twist the lattice action is invariant under 
the further transformation (which does not require changing sign to $r$) 
${\cal{P}} \times {\cal{D}}_d\times (m_q\to -m_q)$, where ${\cal{P}}$ is 
ordinary parity (${x_P=(x_0,-{\bf x})}$)
\begin{eqnarray}
{{\cal {P}}:\left \{\begin{array}{ll}
\hspace{-.5cm}&\psi(x)\rightarrow \gamma_0 \psi(x_P)
\, , \quad \bar{\psi}(x)\rightarrow\bar{\psi}(x_P)\gamma_0\\
\hspace{-.5cm}&U_0(x)\rightarrow U_0(x_P)\, ,\\
\hspace{-.5cm}&U_k(x)\rightarrow U_k^{\dagger}(x_P-a\hat{k})
\, ,\quad k=1,2,3
\end{array}\right .}\label{PAROP}\end{eqnarray}
This fact allows to prove that (for parity-even operators) O($a^{2k+1}$), 
$k\geq 0$, improvement is automatic. The proof (see Appendix of ref.~\cite{FMPR}) 
closely follows the previous line 
of reasoning and does not depend on the $r$-parity of $M_{\rm cr}$. 
One starts with eq.~(\ref{SYMEXP1}) taken at, say, 
$\omega=\pi/2$. Using the lattice symmetry 
${\cal{P}} \times {\cal{D}}_d\times (m_q\to -m_q)$, it yields 
\begin{eqnarray}
\hspace{-.8cm}&&\langle O(x)\rangle\Big{|}_{(r,m_q)}^{\pi/2}=  
(-1)^{P[O]+d[O]}\langle O(-x_P)\rangle\Big{|}^{\pi/2}_{(r,-m_q)}\!\!=\nonumber\\
\hspace{-.8cm}&&=(-1)^{P[O]+d[O]}\Big{[}\zeta^{O}_{O}(r)\langle O(-x_P) \rangle+\nonumber\\
\hspace{-.8cm}&&+a\sum_{\ell}\eta^{O}_{O_{\ell}}(r)
\langle O_{\ell}(-x_P)\rangle\Big{]}^{\rm{cont}}_{(-m_q)}\!\!\!+{\rm{O}}(a^2)
\label{MAXTEXP1}\end{eqnarray}
In eq.~(\ref{MAXTEXP1}) $P[O]$ is the parity of $O$. As before, we 
can act with the corresponding continuum symmetry transformation on the r.h.s.\ of 
this equation to bring back its space-time argument from $-x_P$ to $x$, obtaining
\begin{eqnarray}
\hspace{-.7cm}&&\langle O(x)\rangle\Big{|}_{(r,m_q)}^{\pi/2}= 
\Big{[}\zeta^{O}_{O}(r)\langle O(x) \rangle+\label{MAXTEXP2}\\
\hspace{-.7cm}&&- a\sum_{\ell}\eta^{O}_{O_{\ell}}(r) (-1)^{P[O_\ell]+P[O]}
\langle O_{\ell}(x)\rangle\Big{]}^{\rm{cont}}_{(m_q)}\!\!+{\rm{O}}(a^2)\, .
\nonumber\end{eqnarray}
Comparing now eq.~(\ref{MAXTEXP1}) with eq.~(\ref{MAXTEXP2}) we get
\begin{equation}
-(-1)^{P[O_\ell]+P[O]}\eta^{O}_{O_{\ell}}(r)=\eta^{O}_{O_{\ell}}(r)\, ,
\label{PARSYM}\end{equation}
which tells us that, if $O$ is a parity-even operator ($P[O]=0$ mod(2)), 
all O($a^{2k+1}$) ($k\geq 0$) terms in the Symanzik expansion of 
$\langle O(x)\rangle|_{(r,m_q)}^{\pi/2}$ are multiplied by 
continuum matrix elements of parity-odd operators that hence 
vanish owing to the parity invariance of the continuum theory. 

\section{Non-degenerate quarks}
\label{sec:NONDEG}

In this section we want to show that the nice properties of Mtm-LQCD, 
i.e.\ automatic improvement and positivity of the fermion determinant,
established for the mass degenerate case, can be extended to the 
non-degenerate case.

It was proved in ref.~\cite{FRC} that the action 
\begin{eqnarray}
\hspace{-.7cm}&&S_{{\rm{F}},nd}[\psi,\bar\psi,U]=\label{NONDEG}\\
\hspace{-.7cm}&&=a^4 \sum_x\,\bar\psi(x)[\gamma\widetilde\nabla
-i\gamma_5{ \tau_1} W_{\rm{cr}}
+m_q+{\tau_3}\epsilon_q]\psi(x)\nonumber
\end{eqnarray}
describes the (lattice regularized) gauge interactions of a pair of 
non-degenerate quarks with renormalized masses 
$\hat{m}_q^{(\pm)}\!\!=\!\!Z_P^{-1}m_q\pm Z_S^{-1}\epsilon_q$, 
where $Z_P$ and $Z_S$ are the renormalization constants of the pseudo-scalar 
and scalar quark density, computed in the massless standard Wilson theory. 
These results follow~\cite{FRC} from the analysis of the chiral WTI's 
associated to the action~(\ref{NONDEG}).

We now prove i) automatic improvement and ii)~positivity of the 
determinant of the Dirac--Wilson operator in~(\ref{NONDEG}).

i) Automatic improvement of v.e.v.'s of parity-even operators follows 
from the invariance of~(\ref{NONDEG}) under the transformation 
${\cal{P}}\times{\cal{D}}_d\times(m_q\to -m_q)\times(\epsilon_q\to -\epsilon_q)$
by an argument which exactly parallels the one developed in sect.~\ref{sec:IMPR} 
for the maximally twisted case.

ii) Positivity of 
\begin{equation}
{\mbox{det}}[D_{nd}]={\mbox{det}}[\gamma\widetilde\nabla-
i\gamma_5\tau_1W_{\rm{cr}}+m_q+\tau_3\epsilon_q]\label{DETND1}
\end{equation}
is proved by first noticing that ${\mbox{det}}[D_{nd}]$ is even in $m_q$ and $\epsilon_q$ 
as it follows from the chain of equalities  
\begin{eqnarray}
\hspace{-.7cm}&&{\mbox{det}}[D_{nd}]={\mbox{det}}[\gamma_5e^{i\pi\gamma_5\tau_1/4}D_{nd}
e^{i\pi\gamma_5\tau_1/4}]=\nonumber\\
\hspace{-.7cm}&&={\mbox{det}}[Q_{\rm{cr}}+i\tau_1 m_q+\gamma_5\tau_3\epsilon_q]=\nonumber\\
\hspace{-.7cm}&&={\mbox{det}}\left[\begin{array}{cc}
Q_{\rm{cr}}+\epsilon_q \gamma_5& im_q \\
im_q & Q_{\rm{cr}}-\epsilon_q \gamma_5
\end{array}\right]\label{DETND2}\end{eqnarray}
where we have defined $Q_{\rm{cr}}=
\gamma_5[\gamma\tilde\nabla+W_{\rm{cr}}]=Q_{\rm{cr}}^\dagger$.
To proceed it is convenient to further write 
\begin{eqnarray}
\hspace{-.7cm}&&{\mbox{det}}[D_{nd}]={\mbox{det}}[Q_{\rm{cr}}^2+m_q^2-\epsilon_q^2]
{\mbox{det}}[1+2\epsilon_q B]\, ,\label{DETND3}\\
\hspace{-.7cm}&&B=X\gamma\tilde\nabla X =-B^\dagger\, ,\,
X=(Q_{\rm{cr}}^2+m_q^2-\epsilon_q^2)^{-1/2}\, .\nonumber\label{DETND4}
\end{eqnarray}
If $m_q^2-\epsilon_q^2\!>\!0$, the first factor in the r.h.s.\ 
of~(\ref{DETND3}) is positive. Expanding in $\epsilon_q B$, we get for the second
\begin{eqnarray}
\hspace{-.7cm}&&\log{\mbox{det}}[1+2\epsilon_q B]={\mbox{tr}}[\log(1+2\epsilon_q B)]=\nonumber\\
\hspace{-.7cm}&&=\!\!-\!\!\sum_{k=1}^{\infty}
\!\frac{(-2\epsilon_q)^k}{k}{\mbox{tr}}[B^k]\!=
\!\!-\!\!\sum_{n=1}^{\infty}\!
\frac{(4\epsilon_q^2)^{n}}{2n}{\mbox{tr}}[(\!-B^\dagger \!B)^n]\!=\nonumber\\
\hspace{-.7cm}&&=\frac{1}{2}{\mbox{tr}}[\log(1+4\epsilon_q^2 B^\dagger B)]\geq 0
\, .\label{DETND5}\end{eqnarray}
In the second equality we have used evenness in $\epsilon_q$ and 
the anti-hermiticity of $B$. The non-negativeness of~(\ref{DETND5}) is 
evident as the argument of the last logarithm is an operator with 
a norm certainly not smaller than unit. 
 
We conclude with the following important observation. We have proved 
the positivity (rather the non-negativity) of the determinant~(\ref{DETND2}) 
of the mass non-degenerate Wilson--Dirac operator under the assumption that the bare 
parameters $m_q$ and $\epsilon_q$ satisfy the inequality $m_q^2-\epsilon_q^2>0$. 
For the renormalized masses of any sea doublet this bound implies 
$\frac{Z_P}{Z_S}>\frac{\hat{m}_q^{(+)}-\hat{m}_q^{(-)}}{\hat{m}_q^{(+)}+\hat{m}_q^{(-)}}$. 
Whether or not such a limitation can be a real problem 
for actual simulations is to be seen. We note that in the worst case, 
taking $\hat{m}_q^{(+)}=\hat m_c\sim 1200$ MeV and 
$\hat{m}_q^{(-)}=\hat m_s\sim 100$ MeV, one gets $\frac{Z_P}{Z_S}>0.85$.

\section{Weak matrix elements}
\label{sec:NOMIX}

The complicated pattern of ``wrong chirality'' mixing affecting 
the construction of the renormalizable effective weak Hamiltonian 
operator on the lattice when Wilson fermions are employed~\cite{BMMRT,DGMSTV} 
has up to now prevented a reliable, full-fledged evaluation of the 
phenomenologically important ${\cal{CP}}$-conserving weak matrix 
elements~\footnote{See, however, ref.~\cite{MRTTSS} for some alternative 
way-outs if the Wilson theory is fully O($a$) improved~\cite{HMPRS,REST}.} 
relevant for the evaluation of $B_K$ and the amplitudes appearing 
in the famous $\Delta I=1/2$ rule~\cite{GAS,REV}. 

Mtm-LQCD offers a comparatively simple framework where the problem of 
``wrong chirality'' mixing can be neatly circumvented leading to a cheap 
computational scheme. As an extra bonus automatic improvement, 
which we discussed in the previous sections, is guaranteed~\cite{FR2}.

The idea is to make recourse to a hybrid formulation in which sea quarks 
are introduced as maximally twisted mass non-degenerate pairs, while valence 
quarks are taken as \"Osterwalder--Seiler (OS) flavour singlet fermions.  
This formulation can be made fully local~\cite{GHO} if for each 
valence quark a ghost field is introduced with the purpose of 
canceling the determinant coming form the valence quark integration. 
Provided sea and valence quarks are given the same renormalized masses, 
the sector of the theory where only correlators with no ghost fields 
are considered is unitary, with ``partial quenching'' effects due a 
slightly different sea and valence quarks regularization  
starting only from order $a^2$~\cite{SHGO}.

The key point on which the rest of this section is focused is the fact 
that killing the effects of  ``wrong chirality'' mixing in the lattice 
data of weak matrix elements requires to consider a specific regularization 
of the valence sector of the theory which depends on the particular 
matrix element one is willing to compute. In particular we will see that 
it may be necessary to replicate certain valence flavours assigning 
appropriately tuned signs to the corresponding Wilson terms.  

Notice that in the Mtm-LQCD hybrid formulation we are considering 
parity and flavour breaking effects can only arise from the 
valence sector. The flavour symmetry of valence sector can be 
larger than the direct product of $N_f$ U(1)-factors, depending 
on the choice of the values of the Wilson parameters.  

\subsection{$B_K$ with no mixing}
\label{sec:BK}

In the formal continuum QCD theory with 4 flavours (QCD4) $B_K$ 
is defined by the formula~\footnote{Undefined notations as well 
as further details on the renormalization properties of the different 
regularizations we will discuss below can be found in ref.~\cite{FR2}
and will not be repeated here for lack of space.} 
\begin{eqnarray}
\hspace{-.7cm}&&
\langle \bar{K}_0 |{ \hat{\cal O}_{VV+AA}^{\Delta S=2} }(\mu) | K_0\rangle
= {16 \over 3} M_K^2 F_K^2 {B_K}(\mu)\label{BKDEF}  \\
\hspace{-.7cm}&&
{{\cal O}_{VV+AA}^{\Delta S=2} = (\bar{s} \gamma_\mu d) (\bar{s} \gamma_\mu d)
+ (\bar{s} \gamma_\mu\gamma_5 d) (\bar{s} \gamma_\mu\gamma_5 d)}
\nonumber\end{eqnarray}
and can be extracted e.g.\ from the correlator 
\begin{eqnarray}
\hspace{-.7cm}&&{C_{K{\cal O}K} } = 
\langle \Phi_K(x) { {\cal O}_{VV+AA}^{\Delta S=2}(0)}
\Phi_K(y) \rangle\, ,\label{CORRBK}\\
\hspace{-.7cm}&&\Phi_K=\bar{d}\gamma_5 s\, .
\label{PHIK}\end{eqnarray}
The key observation of ref.~\cite{FR2} is that the same information can be 
extracted from the correlator 
\begin{eqnarray}
\hspace{-.7cm}&&{C_{K'{\cal Q}K}}\!\! =
\langle \Phi_{K'_{v}}(x){2}{{\cal Q}_{VV+AA}^{\Delta S=2}(0)}
\Phi_{K_{v}}(y) \rangle\, , \label{NCOR}\\
\hspace{-.7cm}&&\Phi_{K'_{v}}=\bar{d}_v'\gamma_5 s_v'\, ,
\quad\Phi_{K_{v}}=\bar{d}_v\gamma_5 s_v\, ,\label{PHIKP}\\
\hspace{-.7cm}&&{\cal Q}_{VV+AA}^{\Delta S=2} = \label{QVVAA}\\
\hspace{-.7cm}&&=(\bar{s}_v \gamma_\mu d_v) (\bar{s}_v' \gamma_\mu d_v')
+ (\bar{s}_v \gamma_\mu\gamma_5 d_v) (\bar{s}_v'\gamma_\mu\gamma_5 d_v') 
+ \nonumber\\
\hspace{-.7cm}&&+(\bar{s}_v \gamma_\mu d_v') (\bar{s}_v' \gamma_\mu d_v)
+ (\bar{s}_v \gamma_\mu\gamma_5 d_v') (\bar{s}_v' \gamma_\mu\gamma_5 d_v)\, ,
\nonumber\end{eqnarray}
where $d_v'$ and $s_v'$ are replicas of $d_v$ and $s_v$ valence 
quarks with $m_v^{d',s'} = m_v^{d,s}$. This statement can be proved by noting 
that, if QCD4 and the theory which for short we will call 4s6v 
(meaning 4 sea quarks - $u_s$, $d_s$, $s_s$, $c_s$ - and 6 valence quarks 
- $u_v$, $u_v'$ $d_v$, $d_v'$, $s_v$, $c_v$) are regularized in the same way 
(say {\it \`a la} GW), the two correlators~(\ref{CORRBK}) 
and~(\ref{NCOR}) are equal, simply because they give rise to exactly 
the same Wick contractions. This also implies that the renormalization constant 
of ${\cal Q}_{VV+AA}^{\Delta S=2}$ in 4s6v is equal to that of 
${\cal O}_{VV+AA}^{\Delta S=2}$ in QCD4   
(if the same renormalizalization condition is employed). Similarly 
$\bar{d_v}\gamma_5 s_v$ and $\bar{d}_v'\gamma_5 s_v'$ have the common 
renormalization constant equal to that of $\bar{d}\gamma_5 s$. 

It is not too difficult at this point to prove that, if we take sea and 
valence quarks of the same flavour to have the same renormalized masses 
and we regularize valence quarks as OS fermions having Wilson terms 
with $r_v^d=r_v^s=r_v^{d'}={-}r_v^{s'}$, the matrix element~(\ref{NCOR}) 
is not contaminated by unwanted ``wrong chirality'' mixing effects.
In fact, the peculiar symmetry properties of the regularized action  
prevent the mixing of ${\cal Q}_{VV+AA}^{\Delta S=2}$ with all the 
other dimension 6 operators with ${\Delta S=2}$ that can be constructed 
with valence quarks. 

The action of the regularized theory 4s6v$^L$ constructed as described 
before is invariant (among others) under the following transformations

i) ${\rm Ex}(d_v,d'_v) \times (m_v^d \leftrightarrow m_v^{d'})$

ii) ${\rm Ex}_5(s_v,s'_v) \times (m_v^s \leftrightarrow -m_v^{s'})$

iii) ${\cal C} \times 
[{\rm Ex}(d_v,s_v)\times (m_v^d \leftrightarrow m_v^{s})]\times\\
\phantom{iii)} \times[{\rm Ex}_5(d'_v,s'_v)\times 
(m_v^{d'}\leftrightarrow -m_v^{s'})]$ 

iv) ${\cal P}_{5} \times (M \rightarrow -M)$.\\ 
where~\footnote{To be precise these transformations must be appropriately
extended to the ghost fields associated to each valence quark in order 
to make the full action invariant~\cite{FR2}.} 
($M$ is the whole set of mass parameter)
{{\[
{\rm Ex}(q^{f_1}_v,q^{f_2}_v) : q^{f_1}_v \rightarrow q^{f_2}_v\quad
\bar{q}_v^{f_1} \rightarrow \bar{q}^{f_2}_v 
\]}}
{\[
{{\rm Ex}_5(q^{f_1}_v,q^{f_2}_v)\!:\!\left \{\begin{array}{lll}
q^{f_1}_v \rightarrow \gamma_5 q^{f_2}_v  &
\bar{q}_v^{f_1} \rightarrow -\bar{q}^{f_2}_v \gamma_5  
\\q^{f_2}_v \rightarrow \gamma_5 q_v^{f_1} & 
\bar{q}^{f_2}_v \rightarrow -\bar{q}_v^{f_1} \gamma_5 
\end{array}\right .}
\]}
\[
{{\cal P}_{5}\!:\!\left \{\begin{array}{ll}
\hspace{-.2cm}U \rightarrow {\cal P}[U] &  
\\
\hspace{-.2cm}q_v^f(x)\! \rightarrow \!\gamma_5\gamma_0 q_v^f(x_P)& 
\!\!\bar{q}_v^f(x)\! \rightarrow\! -\bar{q}_v^f(x_P) \gamma_0\gamma_5 \\
\hspace{-.2cm}\psi^j(x)\! \rightarrow\! \gamma_5\gamma_0 \psi^j(x_P)&  
\!\!\bar{\psi}^j(x)\! \rightarrow\! -\bar{\psi}^j(x_P) \gamma_0\gamma_5
\end{array}\right . }
\]
and ${\cal C}$ is charge conjugation.
One checks that the operator ${\cal Q}_{VV+AA}^{\Delta S=2}$ is even 
under any of the above transformation, while the dimension 6 operators 
with which it can potentially mix (namely ${\cal Q}^{\Delta S=2}_{{VV-AA}}$,
${\cal Q}^{\Delta S=2}_{{SS-PP}}$, ${\cal Q}^{\Delta S=2}_{{TT}}$, 
${\cal Q}^{\Delta S=2}_{{SS+PP}}$, ${\cal Q}^{\Delta S=2}_{{VA \pm AV}}$, 
${\cal Q}^{\Delta S=2}_{{SP \pm PS}}$ and 
${\cal Q}^{\Delta S=2}_{{T\widetilde{T} }}$) are all odd. 

\subsection{Kaon decay amplitudes}
\label{sec:KDA}

A long standing puzzle in low energy hadron physics, often 
referred to as ``octet enhancement'' or ``$\Delta I = 1/2$ rule'', 
is represented by the surprisingly large experimental value of 
the ratio~\cite{GAS}
\begin{equation}  
{\rm R}(K\to\pi\pi) = {\Gamma(K\to \pi\pi)\vert_{\Delta
I=1/2}\over\Gamma(K\to \pi\pi)\vert_{\Delta I=3/2}}\,\sim\,
400\, .\label{DELI}\end{equation} 
Though the rate of the $\Delta I=3/2\,K\to\pi\pi$ weak 
decays can be reasonably well computed within our present 
understanding of field theory (based on O.P.E.\ and 
renormalization group arguments), theoretical estimates of the
$\Delta I=1/2$ amplitude give much too small values 
compared to the experimental number~\cite{BB}. 

The lattice approach seems to be the natural framework where a 
{\it first principle} evaluation of such amplitides can be 
carried out. There are, however, severe difficulties in the process of 
establishing a proper strategy to accomplish this task.
First of all, in Euclidean metric the procedure necessary to extract
the kaon decay amplitudes of physical interest 
is complicated by IR subtleties arising from final (two-pion) state
interactions~\cite{MT}. To attack this problem new interesting
ideas have been recently put forward~\cite{MTWAYOUT,LMST}. 
Here we will only deal with the UV
difficulties related to the construction of the renormalized
effective weak Hamiltonian operator on the lattice and show 
how exploiting the flexiblity of the Mtm-LQCD formulation the problem 
of wrong chirality mixing can be brought to an amenable solution.
For an interesting step forward in this direction see also~\cite{PSV}. 

\subsubsection{Generalities}
\label{sec:GEN}

In the Standard Model the decay of the $K$ meson into pions is 
described to leading order in the Fermi constant, $G_{\rm F}$, by the
${\cal{CP}}$-conserving, $\Delta S=1$ effective weak Hamiltonian, which 
(in the chiral limit) reads~\cite{GAS}
\begin{equation}
{\cal{H}}_{\rm{eff}}^{\Delta S=1} = V_{ud}V_{us}^* {G_{\rm F}
\over \sqrt{2}} \sum_{\ell=\pm}C_{\ell}(\mu/M_{\rm W}) \widehat{\cal
O}^{\ell}(\mu) 
\, , \label{HLL} \end{equation}
with $V_{ud}V_{us}^*$ the product
of the appropriate elements of the CKM matrix~\footnote{As
usual, the top quark contribution, which is down by a factor
O($V_{td}V^\star_{ts}/V_{ud}V^\star_{us}\simeq 10^{-3}$), is
neglected.}. The effective operator
${\cal{H}}_{\rm{eff}}^{\Delta S=1}$ is obtained after having
integrated out all degrees of freedom above some energy scale,
$\Lambda$, with $\Lambda$ larger than the charm threshold, but
still well below the W-boson mass, $M_{\rm W}$ Consistently, the
operators $\widehat{\cal O}^{\pm}$ in eq.~(\ref{HLL}) are
renormalized at the scale $\mu$ with $m_c\ll\mu\ll M_{\rm W}$, while 
the Wilson coefficients $C_{\pm}(\mu/M_{\rm W})$ carry the
information about the physics between $\mu$ and $M_{\rm W}$. 
The explicit expression of the bare operators corresponding to 
$\widehat{\cal O}^{\pm}$ is ($\gamma_\mu^L = \gamma_\mu (1-\gamma_5)$)
\begin{eqnarray} 
2{\cal O}^{\pm} \!\!= \!
[(\bar{s}\gamma_\mu^L u) (\bar{u} \gamma_\mu^L d) 
\pm (\bar{s}\gamma_\mu^L d) (\bar{u} \gamma_\mu^L u)]\!-\!
[ u \!\leftrightarrow \!c]\, .\nonumber\end{eqnarray} 
To make contact with experimental data it is
enough to consider the decay of the neutral kaon, $K^0$, into
either $\pi^+ \pi^-$ or $\pi^0\pi^0$ states~\cite{CB,MMRT,HL,PSV}.
Owing to the parity invariance of the formal continuum QCD4 theory,
the relevant amplitudes ${\cal A}({K^0 \to \pi^+\pi^-})$ and 
${\cal A}({K^0 \to \pi^0\pi^0})$
can be written in terms of matrix elements of the renormalized parity
odd operators, $\widehat{\cal O}^{\pm}_{VA}(\mu)$, whose bare 
expression is 
\begin{eqnarray}
\hspace{-.7cm}&&2{\cal O}^{\pm}_{VA}\!=\!
[(\bar{s}\gamma_\mu u) (\bar{u}
\gamma_\mu\gamma_5 d) \pm (\bar{s}\gamma_\mu d) (\bar{u}
\gamma_\mu\gamma_5 u) \! +\nonumber \\
\hspace{-.7cm}&&+(\bar{s}\gamma_\mu\gamma_5 u) (\bar{u} \gamma_\mu
d) \pm (\bar{s}\gamma_\mu\gamma_5 d) (\bar{u} \gamma_\mu u)]
\!-\![ u \leftrightarrow c ] \, .\nonumber
\end{eqnarray}

\subsubsection{$K\to\pi\pi$ with no mixing}
\label{sec:KPIPI}

The idea of the approach developed in ref.~\cite{FR2} is to 
extend the philosophy employed in sect.~\ref{sec:BK} for the computation 
of $B_K$ to the kaon decay amplitudes. In this case the 
required pattern of valence quark replicas will be somewhat more 
complicated and one needs to take four replicas of both $u_v$ 
and $c_v$ valence flavours. We immediately notice that this entails 
only a small extra computational burden (the evaluation of six rather 
than four valence quark propagators). 

More precisely it can be shown that any information contained in 
the QCD4 correlator (meson charges are understood)
\begin{equation}
C_{\pm , K\pi\pi}^{\mbox{\small{QCD4}}}\!
\!=\langle\Phi_{\pi}(x) \Phi_{\pi}(z) {{\cal O}}^{\pm}_{VA}(0)
\Phi^\dagger_{K}(y) \rangle|^{\mbox{\small{QCD4}}}\label{KPPQCD}\end{equation}
can be recovered from the correlator 
\begin{equation}
C_{\pm , K\pi\pi}^{{4{s}10{v}}}\!=\!\langle
\Phi_{\pi}(x) \Phi_{\pi}(z){{\cal Q}}^{\pm}_{VA}(0)
\Phi^\dagger_{K}(y)\rangle|^{{4{s}10{v}}}\, .
\label{KPP4S}\end{equation}
The model ${4{s}10{v}}$ is intended as a theory where, 
besides the two doublets of sea quarks $(u,d)$ and $(s,c)$, in 
the valence sector the $u_v$ and $c_v$ flavours are replicated four 
times. Giving them the names $u_v^{[k]}$ and $c_v^{[k]}$ ($k=0,1,2,3$), 
${{\cal Q}}^{\pm}_{VA}$ 
must be taken to have the expression~\footnote{From 
now on not to have much too clumsy notation we will 
drop the subscript $_v$ on valence quark fields.} 
\begin{eqnarray}
\hspace{-.7cm}&&{{\cal Q}}^{\pm}_{VA} = {\cal Q}^{\pm \, [0]}_{VA}
+{\cal Q}^{\pm \, [1]}_{VA} - {1 \over 2} {\cal Q}^{\pm \, [2]}_{VA}
-{1\over 2} {\cal Q}^{\pm \,[3]}_{VA} \, ,\label{QVA}\\
\hspace{-.7cm}&&{\cal Q}^{\pm\,[k]}_{VA}=
{\cal O}^{\pm}_{VA}|_{u=u^{[k]}}^{c=c^{[k]}}\, .\label{DEFQ}\end{eqnarray}
We remark that the interpolating pion and kaon fields in eq.~(\ref{KPP4S}), 
besides the quarks $d$ and $s$, should be constructed with only the 
$u^{[0]}$ replica of the up quark. 
As in sect.~\ref{sec:BK}, one can prove that, if the two theories QCD4 
and ${4{s}10{v}}$ are regularized in the same way, the 
correlators~(\ref{KPPQCD}) and~(\ref{KPP4S}) are equal. 

At this point the theory ${4{s}10{v}}$ is regularized 
by taking the signs of the Wilson parameters of  
OS valence flavours to be in the particular relation   
\begin{eqnarray}
&&r_d\!=\!{r_s}\!=\!r_{u^{[0]}}\!=
\!{-}r_{u^{[1]}}\!=\!r_{u^{[2]}}\!=\!{-}r_{u^{[3]}}\!=\nonumber\\
&&=\!r_{c^{[0]}}\!=\!{-}r_{c^{[1]}}\!=\!r_{c^{[2]}}\!=\!{-}r_{c^{[3]}}\, . 
\label{RFAC1}\end{eqnarray}
and the renormalized masses of (sea and valence) quarks of the same 
flavour are assigned the same value. We will give the name ${4{s}10{v}}^L$ 
to this regularization of the ${4{s}10{v}}$ theory. 

One can prove that in ${4{s}10{v}}^L$ the operators
${{\cal Q}}^{\pm}_{VA}$ are multiplicatively renormalizable, 
i.e. i)~they mix neither with operators of dimension 
smaller than 6, except for $(m_c^2 - m_u^2)(m_s - m_d)\bar{s}\gamma_5 d$ 
which comes into play with a logarithmic divergent coefficient, 
ii) nor with operators of dimensions 6. 
 
i) We first prove that the pattern of mixing with operator of 
dimension $<6$ is as in the continuum, namely that the operators 
of dimension 3 and 5 [$\bar{s}\gamma_5 d$, $\bar{s}\gamma_5\sigma\!\cdot\!Fd$] 
can only appear multiplied by the mass factors $(m_c^2-m_u^2)(m_s-m_d)$ 
and the operators [$\bar{s}d$, $\bar{s}\sigma\!\cdot\!Fd$] by
$(m_c^2-m_u^2)(m_s^2-m_d^2)$. The various mass factors come from the 
symmetries enjoyed by the regularized theory ${4{s}10{v}}^L$, 
according to the following scheme.

\noindent $\bullet$ $m_c-m_u$ from 
$\prod_k {\rm Ex}(u^{[k]},c^{[k]})\times(m_u \leftrightarrow m_c)$,

\noindent $\bullet$ $m_s-m_d$ from ${\rm Ex}(d,s) \times {\cal C}
\times (m_d \leftrightarrow m_s)$, 

\noindent $\bullet$ $m_c+m_u$ from $[{\rm Ex}_5(u^{[0]},u^{[1]})
\times{\rm Ex}_5(u^{[2]},u^{[3]})\times$\\
\phantom{\noindent $\bullet$} $\times(m_u\rightarrow -m_u)]
\times [u\rightarrow c]$,

\noindent $\bullet$ $m_s-m_d$ from ${\cal P}_5\times(M\rightarrow -M)$, 
but only in\\
\phantom{\noindent $\bullet$} association with the operators $\bar s d$ and  
$\bar{s}\sigma\!\cdot\! F d$.

ii) Proving that the operators ${{\cal Q}}^{\pm}_{VA}$ do not mix with operators 
of dimension 6 can be done by setting all quark masses to zero ($M=0$). 
Following~\cite{FR2}, we proceed by showing that the symmetries of the 
regularized theory forbid the mixing of anyone of the four operators 
${\cal Q}^{\pm\,[k]}_{VA}$ appearing in eq.~(\ref{QVA}) with all the 
other operator of dimension 6 having the same unbroken quantum numbers. 
Let us start by discussing the case of ${\cal Q}^{\pm\,[0]}_{VA}$ and 
${\cal Q}^{\pm\,[2]}_{VA}$. We first notice that the $SU(4)$ flavour 
symmetry of the massless theory prohibits mixing between $+$ and $-$ operators. 
Absence of mixing with operators having tensor structures other than $VA+AV$ 
is proved by identifying the various symmetries that forbid them. The result is  

\noindent $\bullet$ ${\cal P}_5$ forbids $VV\pm AA$, $SS\pm PP$ and $TT$.

\noindent $\bullet$ ${\rm Ex}(d,s)\times{\cal C}$ 
forbids $SP+PS$ and $T\tilde T$.

\noindent $\bullet$ ${\rm Ex}(d,s)\times{\cal C}\times SU(4)_v$ forbids  
$SP-PS$ and\\ \phantom{\noindent $\bullet$} $VA-AV$.\\
The analysis of the mixing properties of ${\cal Q}^{\pm[1]}_{VA}$ 
(${\cal Q}^{\pm[3]}_{VA}$) is immediately brought back to the case of
${\cal Q}^{\pm[0]}_{VA}$ (${\cal Q}^{\pm[2]}_{VA}$) discussed before, 
by the observation that the change of variables induced by ${\cal R}_5$ on the 
valence quarks $u^{[1]}$ and $c^{[1]}$ ($u^{[3]}$ and $c^{[3]}$)
changes sign to $r_{u^{[1]}}$ and $r_{c^{[1]}}$ ($r_{u^{[3]}}$ and 
$r_{c^{[3]}}$), making them equal to $r_{u^{[0]}}$ and $r_{c^{[0]}}$ 
($r_{u^{[2]}}$ and $r_{c^{[2]}}$).

\subsubsection{$K\to\pi$ with no mixing}
\label{sec:KPI}

By the use of soft-pion theorems one can relate $K\to\pi\pi$ to $K\to\pi$ 
amplitudes~\cite{CB,BMMRT} in the chiral limit with the advantage that only 
three-point correlators need be evaluated and no problems with final state 
interactions occur. 

In this approach the relevant target matrix elements are, 
for instance, $\langle\pi^+,{\bf q}|{{\cal O}}^{\pm}_{VV+AA}|K^+,{\bf p}\rangle$. 
They can be extracted from the correlators 
\begin{eqnarray}\hspace{-.7cm}&&C_{\pm , K\pi}^{\mbox{\small{QCD4}}}\!\!=\!
\langle\Phi_{\pi}(x) {{\cal O}}^{\pm}_{VV+AA}(0)
\Phi^\dagger_{K}(y)\rangle|^{\mbox{\small{QCD4}}}\, ,\hspace{-1.cm}\label{KP}\\
\hspace{-.7cm}&&2{{\cal O}}^{\pm}_{VV+AA} \!\!= \![\! (\bar{s}\gamma_\mu u) 
(\bar{u}\gamma_\mu d) \!\pm \!(\bar{s}\gamma_\mu d)(\bar{u}\gamma_\mu u)
\!+\hspace{-1.cm}\label{OVVAA}\\
\hspace{-.7cm}&& +(\bar{s}\gamma_\mu\gamma_5 u) (\bar{u} \gamma_\mu\gamma_5 d) 
\!\pm\! (\bar{s}\gamma_\mu\gamma_5 d) (\bar{u} \gamma_\mu \gamma_5u)\! ]
\!\!-\!\![u\!\leftrightarrow\! c ]\, .
\nonumber\end{eqnarray}
Noting that, if QCD4 and $4{s}10{v}$ are regularized in the 
same way, the Wick theorem ensures the equality of all contractions, 
we conclude that the same information is contained in 
\begin{eqnarray}
C_{\pm , K\pi}^{{4{s}10{v}}}= \langle
\Phi_{\pi}(x) { {\cal Q}}^{\pm}_{VV+AA}(0)
\Phi^\dagger_{K}(y) \rangle|^{4{s}10{v}}\, ,
\label{NCORR}\end{eqnarray}
As before, there is a suitable regularization of $4{s}10{v}$ which 
makes the operators ${\cal Q}^{\pm}_{VV+AA}$ multiplicatively 
renormalizable. Besides taking all the renormalized quark masses 
of the same flavour equal, one needs to choose the 
Wilson parameters of the OS valence quarks obeying the relation 
\begin{eqnarray}
&&r_d\!=\!-{r_s}\!=\!r_{u^{[0]}}\!=
\!{-}r_{u^{[1]}}\!=\!r_{u^{[2]}}\!=\!{-}r_{u^{[3]}}\!=\nonumber\\
&&=\!r_{c^{[0]}}\!=\!{-}r_{c^{[1]}}\!=\!r_{c^{[2]}}\!=\!{-}r_{c^{[3]}}\, . 
\label{RFAC2}\end{eqnarray}
We notice that the only difference with respect to the choice~(\ref{RFAC1}) 
is the sign change in front of $r_s$. We will call this regularization 
$4{s}10{v}^{L\star}$. 

The proof of this statement follows from the invariance of expectation
values under the change of fermionic integration variables induced by the 
transformation $s \rightarrow  s'=\gamma_5 s$, 
$\bar{s} \rightarrow  \bar{s}'=-\bar{s}\gamma_5$ on the valence strange 
quark (the associated ghost fields should be simultaneously 
transformed in the appropriate way~\cite{FR2}). Under this change of 
variables the action of $4{s}10{v}^{L\star}$ goes over to the action 
of the $4{s}10{v}^{L}$ theory of sect.~\ref{sec:KPIPI},
modulo the change of sign of the valence quark mass, $m_s$. 
At the same time ${\cal Q}^{\pm}_{VV+AA}\rightarrow{\cal Q}^{\pm}_{VA}$ 
and $(m_s+m_d)\bar sd\rightarrow (-m_s+m_d)\bar s\gamma_5 d$. We 
are thus exactly in the situation discussed in sect.~\ref{sec:KPIPI}, 
except that now the operators [$\bar{s}d$, $\bar{s}\sigma\!\cdot\!Fd$] 
will appear multiplied by the mass factors $(m_c^2-m_u^2)(m_s+m_d)$ 
and the operators [$\bar{s}\gamma_5d$, $\bar{s}\gamma_5\sigma\!\cdot\!Fd$] 
by $(m_c^2-m_u^2)(m_s^2-m_d^2)$. 

This argument also proves that the operators 
${\cal O}^{\pm}_{VV+AA}$ in QCD4 and ${\cal Q}^{\pm}_{VV+AA}$ in $4{s}10v^{L*}$ 
(can be taken to) have equal renormalization constant and the same as 
${\cal Q}^{\pm}_{VA}$ in $4{s}10{v}^{L}$. 

We conclude by stressing that in all the applications considered in 
sect.~\ref{sec:NOMIX}, full O($a^{2k+1}$), $k\geq 0$, improvement is 
ensured by the symmetry of the regularized actions under 
${\cal P}\times{\cal D}_d\times(M\rightarrow -M)$.

\section{Chirally enhanced cutoff effects}
\label{sec:OPTCH}

Although, as we have seen, in Mtm-LQCD odd power discretization effects   
are absent or easily eliminated~\cite{TM,FR1,FRC,FR2}, it turns out that correlators 
are affected by dangerous artifacts of relative order $a^{2k}$, $k\geq 1$, 
which are enhanced by inverse powers of the (squared) pion mass, as the latter 
becomes small~\cite{FMPR}. In fact, when analyzed in terms of the Symanzik 
expansion, lattice expectation values exhibit, as $m_\pi^2\to 0$, what we will 
call ``infrared (IR) divergent'' cutoff effects with a behaviour of the form 
($2k\geq h\geq 1$, $k,h\,\,{\rm integers}$)
\begin{equation}
\<O\>\Big{|}^L_{m_q}=\<O\>\Big{|}^{\rm cont}_{m_q}
\Big{[}1+{\rm O}\Big{(}\frac{a^{2k}}{(m_\pi^2)^{h}}\Big{)}\Big{]}
\, ,\label{ORD}\end{equation}
where we have assumed that the lattice correlator admits a 
non-trivial continuum limit. Powers of $\Lambda_{\rm QCD}$ required 
to match physical dimensions are often understood in this section. 

We want to prove that artifacts of the type~(\ref{ORD}) are reduced to terms 
that are at worst of order $a^{2}(a^2/m_\pi^2)^{k-1}$, $k\geq 1$, 
if the action is O($a$) improved {\it \`a la} Symanzik or, alternatively,
if the critical mass is chosen in some ``optimal'' way.

The idea that a suitable definition of critical mass exists which 
can lead to a smoothing out of chirally enhanced lattice artifacts or
perhaps be of help in getting improvement was already put forward in the 
context of lattice $\chi$PT in refs.~\cite{SHWUNEW} and~\cite{AB}, respectively.

An important consequence of this analysis is that the strong (order of 
magnitude) inequality $m_q> a\Lambda^2_{\rm QCD}$, invoked in ref.~\cite{FR1}, 
can be relaxed to the weaker relation $m_q> a^2\Lambda^3_{\rm QCD}$,
before large cutoff effects are possibly met while lowering the quark mass 
at fixed $a$. The works of refs.~\cite{SHWUNEW,AB}, and most recently
refs.~\cite{AB05,SH05}, all based on lattice $\chi$PT, lead to 
essentially equivalent conclusions about cutoff effects in pion quantities 
in the parameter region $m_q> a^2 \Lambda^3_{\rm QCD}$. They also yield 
interesting predictions on the possible Wilson fermion phase 
scenarios~\cite{META1,META2} and observables, when $m_q$ is of order $a^2$ or smaller. 

A nice numerical demonstration of the effectiveness of Mtm-LQCD in killing O($a$) 
discretization errors and the ability of the optimal choice of the critical 
mass in diminishing the magnitude of lattice artifacts at small quark mass 
can be found in refs.~\cite{CAN,XLFNEW,SHI}. 
As for Mtm-LQCD with clover-improved quark action, the promising quenched tests 
presented some years ago in~\cite{DMETAL} have been recently extended 
in~\cite{LUB05} down to pion masses of 300~MeV or lower, confirming the 
absence of large cutoff effects. 

\subsection{Symanzik analysis of ``IR divergent'' cutoff artifacts}
\label{sec:SEOLC}  

The study of discretization artifacts affecting lattice correlators 
in Mtm-LQCD can be elegantly made in the language of the Symanzik 
expansion. We focus here on Mtm-LQCD with two mass degenerate flavours.
Its fermionic action is given by eq.~(\ref{STM}) with $\omega=\pi/2$. 
A full analysis of cutoff effects beyond O($a$) is extremely 
complicated. Fortunately it is not necessary, if we limit the discussion 
to the terms that are enhanced as the quark mass $m_q$ is decreased.

$\bullet$ {\it The Symanzik LEEA of Mtm-LQCD} - 
The low energy effective action (LEEA), $S_{\rm Sym}$, of 
the theory can be written in the form
\begin{eqnarray}
S_{\rm Sym}=\int\!d^4y\,\Big{[}{\cal L}_4(y)+
\sum_{j=0}^{\infty}a^{j}\ell_{4+j}(y)\Big{]}\, ,
\label{SLEEA}\end{eqnarray}
where ${\cal L}_4=\frac{1}{2g_0^2}{\rm tr}(F\!\cdot\! F)+ 
\bar\psi(\gamma \!\cdot\! D + m_q)\,\psi$ is the target 
continuum 2-flavour QCD Lagrangian. Based on the symmetries
of Mtm-LQCD a number of interesting properties enjoyed 
by $S_{\rm Sym}$ can be proved which are summarized below.

1. Lagrangian densities of even dimension, $\ell_{4+2k}$, in 
eq.~(\ref{SLEEA}) are parity-even, while terms of odd dimension, 
$\ell_{5+2k}$, are parity-odd and twisted in iso-spin space. 
Thus the latter have the quantum numbers of the neutral pion. 

2. The term of order $a$ in eq.~(\ref{DEFEO}), $\ell_5$, 
is given (on-shell) by the linear combination
\begin{eqnarray}
\hspace{-.7cm}&&\ell_5=\delta_{5,SW}\,\ell_{5,SW}+\delta_{5,m^2}\,
\ell_{5,m^2}+\delta_{5,e}\,\ell_{5,e}\, ,\label{L5}\\
\hspace{-.7cm}&&\ell_{5,SW}=
\frac{i}{4}\bar\psi[\sigma\cdot F]i\gamma_5\tau_3\psi\, ,
\!\!\quad\ell_{5,m^2} = m_q^2 \bar\psi i\gamma_5\tau_3\psi\, ,\nonumber\\
\hspace{-.7cm}&&\ell_{5,e}=\Lambda_{\rm QCD}^2\bar\psi i\gamma_5\tau_3\psi\, ,
\label{L51}
\end{eqnarray}
where $\delta_{5,SW}$, $\delta_{5,m^2}$ and $\delta_{5,e}$ are 
dimensionless coefficients, odd in $r$. The operator $\ell_{5,e}$ 
arises from the need to describe order $a$ uncertainties entering any 
non-perturbative determination of the critical mass and goes together 
with $\ell_{5,SW}$. Both $\ell_{5,SW}$ and $\ell_{5,e}$ could be made to 
disappear from~(\ref{SLEEA}) by introducing in the Mtm-LQCD action the 
SW (clover)-term~\cite{SW} with the appropriate non-perturbatively 
determined $c_{SW}$ coefficient~\cite{LU} and at the same time setting 
the critical mass to its correspondingly O($a$) improved value. 

3. Higher order ambiguities in the critical mass, which 
will all contribute to ${\cal L}_{\rm odd}$, are described by terms 
proportional to odd powers of $a$ of the kind 
\begin{eqnarray}
\hspace{-.7cm}&&a^{2k+1}\,\delta_{5+2k,e}\,\ell_{5+2k,e}=\nonumber\\
\hspace{-.7cm}&&=\!a^{2k+1}\delta_{5+2k,e}\,
(\Lambda_{\rm QCD})^{2k+2}\bar\psi i\gamma_5\tau_3\psi\, ,\,\,k\geq 1\, .
\label{HK}\end{eqnarray}

$\bullet$ {\it Describing lattice correlators beyond O($a$)} - 
We are interested in the Symanzik description of the lattice artifacts 
affecting the correlator $\<O\>|^L_{m_q}$, where $O$ has 
continuum vacuum quantum numbers so as to yield a non trivially vanishing 
result as $a\to 0$. Schematically we write 
\begin{eqnarray}
\hspace{-.7cm}&&\<O({x})\>\Big{|}_{m_q}^{L}\!=\!
\<[O({x})+\Delta_{\rm odd}O(x)+\Delta_{\rm even}O(x)]\nonumber\\
\hspace{-.7cm}&&e^{-\int\!d^4 y[{\cal L}_{\rm odd}(y)+{\cal L}_{\rm even}(y)]}\>
\Big{|}^{\rm cont}_{m_q}\, , \label{SEOP}\\
\hspace{-.7cm}&&{\cal L}_{\rm odd}\!=\!\sum_{k=0}^{\infty}a^{2k+1}\ell_{5+2k}\, ,
{\cal L}_{\rm even}\!=\!\sum_{k=1}^{\infty}a^{2k}\ell_{4+2k}\, .
\label{DEFEO}\end{eqnarray}
The operators $\Delta_{\rm odd}O$ ($\Delta_{\rm even}O$)
have an expansion in odd (even) powers of $a$. They can be viewed as the 
operators necessary for the improvement of the 
matrix elements of $O$~\cite{HMPRS,LU}. To ensure automatic improvement~\cite{FR1} 
we shall assume that $O$ is parity invariant in which case its Symanzik 
expansion will contain only even powers of $a$. 

$\bullet$ {\it Pion poles and ``IR divergent'' cutoff effects} - 
Although a complete analysis of all the ``IR divergent'' cutoff effects 
is very complicated, the structure of the leading ones ($h=2k$ in 
eq.~(\ref{ORD})) is rather simple, as they only come from continuum 
correlators where $2k$ factors $\int d^4y {\cal L}_{\rm odd}(y)$ 
are inserted. More precisely the leading ``IR divergent'' cutoff 
effects are identified on the basis of the following result~\cite{FMPR}. 

In the Symanzik expansion of $\<O\>|^L_{m_q}$ at order $a^{2k}$ 
($k\geq 1$) there appear terms with a $2k$-fold pion pole and residues 
proportional to $|\<\Omega|{\cal L}_{\rm odd}|\pi^0({\bf 0})\>|^{2k}$, 
where $\<\Omega|$ and $|\pi^0({\bf 0})\>$ denote the vacuum 
and the one-$\pi^0$ state at zero three-momentum, respectively. 
Putting different factors together, each one of these terms can be seen 
to be schematically of the form (${\cal L}_{\rm odd}={\rm O}(a)$) 
\begin{eqnarray}
\hspace{-.7cm}&&\Big{[}\Big(\frac{1}{m_\pi^2}\Big)^{2k}
(\xi_{\pi})^{2k}{\cal M}[O;\{\pi^0({\bf 0})\}_{2k}]
\Big{]}_{m_q}^{\rm cont}\, ,\label{METREO}\\
\hspace{-.7cm}&&\xi_{\pi}=|\<\Omega|{\cal L}_{\rm odd}|
\pi^0({\bf 0})\>|_{m_q}^{\rm cont}\, ,\label{XSI}\end{eqnarray}
where we have generically denoted by ${\cal M}[O;\{\pi^0({\bf 0})\}_{2k}]$ 
the $2k$-particle matrix elements of $O$, with each external leg 
being a zero three-momentum neutral pion. 

Less ``IR divergent'' cutoff effects (those with $h$ strictly 
smaller than $2k$ in~(\ref{ORD})) come either from terms 
with some extra $\int d^4y{\cal L}_{\rm even}(y)$ insertions or from
contributions of more complicated intermediate states other than 
straight zero three-momentum pions or from both. In all these cases 
one gets extra $a^2$ powers, not all ``accompanied'' by corresponding 
$1/(m_\pi^2)^2$ factors.

It is important to stress that the appearance of pion poles like the 
ones in~(\ref{METREO}) in no way means that the lattice correlators 
diverge as $m_q\to 0$, but only that the Symanzik expansion we have 
employed appears to have a finite radius of convergence (on this point 
see the remarks in~\cite{SH05}).

\subsection{Reducing ``IR divergent'' cutoff effects}
\label{sec:KLL}

Recalling that ${\cal{L}}_{\rm odd}=a\ell_5+{\rm O}(a^3)$, the previous
analysis shows that at leading order in $a$ the residue of the most 
severe multiple pion poles is proportional to 
$|\<\Omega|\ell_{5}|\pi^0({\bf 0})\>|^{2k}$. It is an immediate conclusion 
then that the leading ``IR divergent'' cutoff effects can all be eliminated 
from lattice data if we can either reduce $\ell_{5}$ to only the ${\ell}_{5,m^2}$ 
term in~(\ref{L5}) or set $\xi_\pi$ to zero. 

$\bullet$ {\it Improving the Mtm-LQCD action} - 
The obvious, field-theoretical way to achieve the first goal 
consists in adding the clover term~\cite{SW,LU,HMPRS} to the Mtm-LQCD action. 
In this case lattice correlators will admit a Symanzik description 
in terms of a LEEA where the operators ${\ell}_{5,SW}$ and 
${\ell}_{5,e}$ are absent, and in $\ell_5$ only ${\ell}_{5,m^2}$ survives. 
The left-over contributions arising from the insertions of ${\ell}_{5,m^2}$ 
yield terms that are at most of order 
$(am_q^2/m_\pi^2)^{2k}\simeq (a m_q)^{2k}$, hence negligible in the chiral 
limit. It is instead the next odd operator in the expansion~(\ref{SLEEA}), 
$a^3\ell_7$, which comes into play. 

A combinatoric analysis based on the structure of the non-leading 
``IR divergent'' cutoff effects reveals that the worst lattice 
artifacts left behind in correlators after the ``clover cure'' are 
of the kind $a^2(a^2/m_\pi^2)^{k-1}$, $k\geq 1$. 

$\bullet$ {\it Optimal choice of the critical mass} - 
The alternative strategy to kill the leading ``IR divergent'' cutoff 
effects consists in leaving the Mtm-LQCD action unimproved,  
but fixing the critical mass through the condition
\begin{equation}
\lim_{m_q \to 0} \!\xi_\pi(m_q)\!=\!\!\lim_{m_q \to 0}
|\<\Omega|{\cal{L}}_{\rm odd}|\pi^0({\bf 0})\>|^{\rm cont}_{m_q}\!=\! 0 \, .
\label{EFFCOND}\end{equation} 
The meaning of~(\ref{EFFCOND}) is simple. It amounts 
to fix, for $k\geq 0$, the order $a^{2k+1}$ contribution in the 
counter-term, $M_{\rm cr}\bar\psi^L i\gamma_5\tau_3\psi^L$, 
so that its vacuum to one-$\pi^0({\bf 0})$ matrix element  
compensates, in the limit $m_q\to 0$, the similar matrix element 
of the sum of all the other operators making up $\ell_{5+2k}$.

Concrete procedures designed to implement condition~(\ref{EFFCOND}) 
in practice were discussed in~\cite{FMPR}. They are all based on the idea of
determining the critical mass by requiring the lattice correlator  
$a^3\sum_{\bf x}\<V_0^2(x)P^1(0)\>|^L_{m_q}$ ($x_0 \neq 0$)
to vanish in the chiral limit ($V_0^2=\bar\psi\gamma_0\frac{\tau_2}{2}\psi$ 
is the vector current with iso-spin index 2 and 
$P^1=\bar\psi\gamma_5\frac{\tau_1}{2}\psi$ the pseudo-scalar density 
with iso-spin index 1). In the continuum this correlator is zero by 
parity for any value of $m_q$. On the lattice the breaking of parity 
(and iso-spin) due to the twisting of the Wilson term makes it non-vanishing 
by O($a$) discretization artifacts, which have the form of a power series 
expansion in $\xi_\pi$. 
 
The important conclusion of the analysis presented in~\cite{FMPR} is 
that it is not necessary (nor possible) to really go to $m_q\to 0$. 
It is enough to have the critical mass determined by the vanishing of
the previous correlator at the smallest available $m_q$-value, say
$m_q^{\rm min}$, provided it fulfills the order of magnitude inequality 
$m_q^{\rm min} > a^2$. Under this condition, for any $m_q \geq m_q^{\rm min}$
we will have $\xi_\pi(m_q)={\rm O}(am_\pi^2)$.   
Substituting this estimate into the leading ``IR divergent'' terms, we 
see that they are reduced to finite O($a^{2k}$) terms. As for the 
sub-leading terms, a non-trivial diagrammatic analysis shows that the worst 
of them, left behind after the ``optimal critical mass cure'', 
are reduced to only $a^2(a^2/m_\pi^2)^{k-1}$, $k\geq 1$, effects, just like 
in the case where the clover term is employed.

The results reached in this section show that the continuum 
extrapolation of lattice data should be smooth at least down to values 
of the quark mass satisfying the order of magnitude inequality 
$m_q >a^2\Lambda^3_{\rm QCD}$, if the clover term is added to the 
lattice action, as well as if the critical mass is adjusted to its ``optimal'' value.

\subsection{The lattice GMOR relation}
\label{sec:GOR}

A workable way to numerically estimate the minimal value of the quark 
mass that can be safely simulated at fixed $a$ can be obtained by 
considering the behaviour of the charged pion mass as a function of $m_q$. 
It turns out, in fact, that in Mtm-LQCD there are Ward-Takahashi identities 
(WTI's) which take exactly the form they have in the formal continuum theory. 
{}From them a lattice GMOR relation can be derived.

To see how this works we recall that in Mtm-LQCD the 1-point split axial 
currents, $\hat{A}_\mu^b$, with iso-spin index $b=1,2$ are exactly conserved 
at $m_q=0$~\cite{TM,FR1}. This implies the validity of the WTI's
($\tau^\pm=(\tau_1\pm i\tau_2)/2$)
\begin{equation}
\<[\partial_\mu^*\hat A_\mu^\pm\!-2m_q P^\pm](x) 
P^\mp(0)\>\Big{|}^L_{m_q}\!\!\!\!=\! 
\< S^0(0)\>\Big{|}^L_{m_q}\!\!\!\delta_{x\!,0}
\label{PCAC}\end{equation} 
\begin{equation}
\hat{A}_\mu^\pm\!\!=\!\hat{A}_\mu^1\pm i\hat{A}_\mu^2\, ,\,
P^\pm\!\!=\!\bar\psi^L\gamma_5\tau^\pm\psi^L\, ,\,
S^0 \!\!=\!\bar\psi^L\psi^L .
\label{P1S0}\end{equation}
After space-time integration, one gets ($m_q\neq 0$)
\begin{equation}
2m_q\, a^4\sum_{x}\< P^\pm(x)P^\mp(0)\>|^L_{m_q}=-\< S^0(0)\>|^L_{m_q} \, .
\label{PCACI}\end{equation}
Although, as the WTI~(\ref{PCACI}) itself shows, there is no mixing between 
$S^0$ and the identity operator with a cubically (or a linearly) divergent 
coefficient, there is still room for a quadratically divergent term proportional 
to $m_q$. Indeed, the l.h.s.\ of~(\ref{PCACI}) can be written as the sum of a piece 
where intermediate states are inserted plus a divergent 
contribution of the kind $m_q/a^2$ coming from the (integrated) short-distance 
singularity of $\<P^\pm(x)P^\mp(0)\>$ at $x\!=\!0$. 
This term should be brought to the r.h.s., thus  
leading to the subtracted expression of the chiral condensate.

We can now repeat on the lattice the argument that in the continuum 
leads to the classical GMOR relation. We will work under the assumption
that spontaneous chiral symmetry breaking occurs in the limiting continuum 
theory, i.e.\ under the assumption that 
\begin{equation}
\Sigma \equiv - \<\Omega|S^0|\Omega\>|^{\rm cont}_{m_q=0}\neq 0\, .
\label{SCSB}\end{equation}
After inserting a complete set of states in the l.h.s.\ of eq.~(\ref{PCACI}),  
we get
\begin{equation}
m^2_{\pi^\pm}\Big{|}^L_{m_q}=
{2m_q}\frac{|\<\Omega|P^\pm|\pi^\pm\>|^2}{[-\<\Omega|S_{\rm sub}^0|\Omega\>]}
\Big{|}^L_{m_q} + \ldots\, ,\label{PCACIR}\end{equation}
where we have explicitely written only the contribution coming from the pion 
pole. Dots are terms due to the intermediate states that stay massive 
as $m_q \to 0$, as well as terms vanishing with $m_q$ faster than linearly.
The subscript ``$_{\rm sub}$'' is to remind us that it is the properly 
subtracted chiral condensate that enters this equation. 
Notice that, as expected, the the r.h.s.\ of eq.~(\ref{PCACIR}) 
is a finite, renormalized, quantity in the limit $a\to 0$.

Once the leading ``IR divergent'' cutoff effects have been canceled out,
the use of the Symanzik expansion in the r.h.s.\ of eq.~(\ref{PCACIR}) yields 
\begin{eqnarray}
m^2_{\pi^\pm}\Big{|}^L_{m_q}\!\!\!\!\!=\!
{2m_q}\frac{G_\pi^2}{\Sigma}
\!\Big{[}\!1\!+\!a^2\!\sum_{\ell\geq 0}\!b_\ell
\Big{(}\frac{a^2}{m_q}\Big{)}^\ell\Big{]}^{\rm cont}\!\!\!+\!\ldots
\, ,\label{PCASYM}\end{eqnarray}
where $G_\pi=|\<\Omega|P^\pm|\pi^\pm\>|^{\rm cont}_{m_q=0}$. 
In eq.~(\ref{PCASYM}) dots denote less dangerous ``IR divergent'' lattice artifacts 
compared to those explicitely shown, as well as contributions of higher 
order in $m_q$. In getting eq.~(\ref{PCASYM}) we have used the fact 
that the continuum limit of $-\<\Omega|S_{\rm sub}^0|\Omega\>^L$  
at vanishing quark mass is $\Sigma\neq 0$ (see eq.~(\ref{SCSB})).

{}From the above analysis it follows that, in the region where 
the series in eq.~(\ref{PCASYM}) converges (i.e.\ at least down to masses 
where the order of magnitude inequality $m_\pi^2\sim m_q >a^2$ is 
still satisfied), the squared mass of the charged lattice pion is 
linear in $m_q$ (up to very small deviations). Thus, vice-versa, 
we can imagine to use deviations from the established linear behaviour 
that are possibly seen at small $m_q$ as a workable criterion to determine the 
minimal value of $m_q$ at which simulations can be performed before 
being set-off by discretization effects. 

\subsection{Artifacts on hadronic energies and $f_\pi$}
\label{sec:ART}

We wish to conclude by discussing some peculiar features concerning 
the magnitude of the O($a^2$) cutoff effects on hadron energies 
(in particular masses) and on the pion decay constant. 

\subsubsection{Hadron energies}
\label{sec:HE}

In the language of the Symanzik expansion discretization artifacts on 
hadronic energies are described by a set of diagrams where at least one
among the inserted $\int {\cal L}_{\rm odd}$ factors gets 
absorbed in a multi-particle irreducible matrix element, with the consequence 
that it is not available for producing a pion pole. As a consequence, at fixed 
order in $a$, the most ``IR divergent'' lattice corrections to continuum 
hadronic energies contain one overall $1/m_\pi^2$ factor less than the 
leading ``IR divergent'' cutoff effects generically affecting correlators.
For instance, at order $a^2$ the difference between lattice
and continuum energy of the hadron $\alpha_n$ reads~\cite{FMPR}
(the label ${\bf q}$ specifying the three-momentum of the state 
$|{\alpha_n}({\bf q})\>$ is omitted) 
\begin{eqnarray}
\hspace{-0.7cm}&&\Delta E_{\alpha_n}({\bf q})\Big{|}_{a^2}\!\propto
\label{DE2LEAD}\\
\hspace{-0.7cm}&&\frac{a^2}{m_\pi^2}
{\rm Re}\frac{\<\Omega| \ell_5 | \pi^0({\bf 0}) \>
\<\pi^0({\bf 0})\alpha_n| \ell_5 |\alpha_n\>}
{2 E_{\alpha_n}({\bf q})}\!+\!{\rm O}(a^2)\Big{|}_{m_q}^{\rm cont}\, ,\nonumber
\end{eqnarray}
where ${\rm O}(a^2)$ denotes ``IR finite'' corrections. 
It should be noted that $\Delta E_{\alpha_n}({\bf q})|_{a^2}$ 
is reduced to a plain  ${\rm O}(a^2)$ ``IR finite'' correction after anyone 
of the two ``cures'' described in Sect.~\ref{sec:KLL}. 
                                                                                        
Specializing~(\ref{DE2LEAD}) to the case of pions, one
obtains the interesting result that the difference between charged
and neutral pion (square) masses is a finite O($a^2$) quantity even
if the critical mass has not been set to its optimal value or the
clover term has not been introduced. The reason is that the
leading ``IR divergent'' contributions shown in~(\ref{DE2LEAD})
are equal for all pions (as one can prove by standard soft pion
theorems~\cite{SPT}), hence cancel in the (square) mass difference.
This conclusion is in agreement with detailed results from chiral 
perturbation theory (see refs.~\cite{CPT} and~\cite{SHWUNEW}),
as well as with the first numerical estimates of the pion
square mass splitting in Mtm-LQCD~\cite{CHRMI}.

\subsubsection{Pion decay constant}
\label{sec:PDC}

Data on a (quenched) computation of $f_\pi$, carried out
in Mtm-LQCD by using the value of the critical mass obtained from
the vanishing of the pion mass, show cutoff effects which tend to
increase as the quark mass is lowered~\cite{KSUW}.
In particular data lie lower than the straight line 
extrapolation drawn from larger masses. This behaviour (called 
``bending phenomenon'' in ref.~\cite{KSUW}) sets in at values 
of the bare quark mass around $a \Lambda_{\rm QCD}^2$, 
$\Lambda_{\rm QCD}\sim 200$~MeV. A recent scaling 
test of Mtm-LQCD~\cite{XLFNEW} furthermore shows that the 
``bending phenomenon'' is an O($a^2$) cutoff effect with a magnitude 
which increases as $m_{\pi^\pm}^2$ is lowered. 

In refs.~\cite{KSUW,CAN,XLFNEW} the lattice pion decay constant, 
$f_\pi^L$, was determined using the formula~\cite{TM,FR1}
\begin{equation}
f_\pi^L(m_{\pi^\pm}^2)=2m_q\frac{\<\Omega|P^\pm|\pi^\mp\>}
{m_{\pi^\pm}^2}\Big{|}^L_{m_q}\, ,\label{FPIL}\end{equation}
from which $f_\pi^L$ is seen to be the ratio of two lattice quantities.
It should be noted that the matrix element in the numerator of 
eq.~(\ref{FPIL}) is affected by leading ``IR divergent'' lattice artifacts  
which are softened by an extra $m_{\pi}^2\sim m_q$ factor. 
This property can be traced back to the invariance of the correlator 
$\< P^\pm(x) P^\mp(y) \>|_{m_q}^L$ 
under axial rotations around the third iso-spin direction. One can, in fact, 
always think of having chosen the -- O($a$) -- rotation angle such to bring 
the critical mass to its optimal value, at the price of shifting the bare 
quark mass by a term of order $a^2/m_q$.  
Also $(m_{\pi^\pm}^L)^2$ is affected by leading ``IR divergent'' cutoff 
effects softened by an extra $m_{\pi}^2$ factor. The latter, however, drops 
out in the ratio $(m_{\pi^\pm}^L)^2/m_q$ entering eq.~(\ref{FPIL}).

With anyone of the two cures described in sect.~\ref{sec:KLL} 
lattice artifacts are reduced to terms that are only of the kind 
$a^2\sum_{\ell\geq 0}c_\ell({a^2}/{m_q})^\ell$~\cite{FMPR}.

Indeed, a beautiful confirmation of the validity of the analysis of 
``IR divergent'' cutoff effects presented here comes from 
the fact that, when the critical mass is set at its optimal value, 
no bending effect is anymore visible in the $f_\pi^L$ data
obtained in ref.~\cite{CAN,XLFNEW} (see also~\cite{SHI}).

\section{Conclusions and outlook}
\label{sec:CONCL}

In this lecture we have reviewed the main properties of Mtm-LQCD 
with the purpose of underlining the features that makes it 
an appealing regularization for realistic simulations of QCD4 
at adequately small pion masses. Actually explorative unquenched 
simulations of QCD4 with renormalized masses satisfying 
$m_u\!=\!m_d\!<\!m_s\!<m_c$ have already started~\cite{NF211}.

We have not discussed the very important issue of 
meta-stabilities that on too coarse lattices appear to affect 
unquenched simulations  and prevent reaching sufficiently light 
pions~\cite{FARC,NF211}, at least in the Singleton--Sharpe 
scenario~\cite{META2}.

In the language of lattice $\chi$PT the strength of meta-stabilities and 
the magnitude of the attainable minimal pion mass are features   
that are both controlled by the magnitude of the coefficient (called $c_2$ 
in refs.~\cite{META2,CPT}) which in the chiral Lagrangian multiplies the 
term $({\mbox{Tr}}[\Sigma+\Sigma^\dagger])^2$ describing O($a^2$) lattice 
artifacts in correlators. Ideally, one thus would like to have $c_2=0$ as a 
function of $g_0^2$. 
This situation is not peculiar of Mtm-LQCD. Similar features arise for 
generic values of the twisting angle and even for plain (clover) Wilson 
fermions a non-vanishing $c_2$ might turn out to be problematic for 
simulations at fixed $a$ if too small pion masses are taken~\cite{META2}.  

Comparing $\chi$PT analysis with results from the Symanzik LEEA approach, 
one can prove that $c_2$ is proportional to the matrix element 
$a^2\<\pi^0({\bf 0})|\ell_6^{\chi{\rm{-br}}}|\pi^0({\bf 0})\>$, where 
$\ell_6^{\chi{\rm{-br}}}$ is the chirally breaking piece of $\ell_6$ 
(see eq.~(\ref{DEFEO})). This observation suggests that a possible way to 
enforce the condition $c_2=0$ is to modify the gauge action by adding 
to the standard plaquette term further dimension 6 operators.
For instance, one can imagine adding $b(g_0^2)P_{2\times 1}$,
where $b(g_0^2)$ is an adjustable coefficient and $P_{2\times 1}$ 
the minimal rectangular plaquette. This term, through the mixing 
induced by the chirally breaking twisted Wilson term, will modify, 
among others, all dimension 6 operators of the Symanzik LEEA, including 
those that break chiral invariance. 
Since $c_2$ also controls the square mass difference between charged and 
neutral pion~\cite{CPT,SHWUNEW,AB}, 
$\Delta m^2_\pi=m^2_{\pi^\pm}-m^2_{\pi^0}\propto c_2$, the idea 
to ease Mtm-LQCD simulations is then to fix $b(g_0^2)$ as a function of $g_0^2$ 
by imposing the condition $\Delta m^2_\pi=0$. 

\vspace{.2cm}
{\bf Acknowledgments --} G.C.R. thanks the Organizers 
for the exciting atmosphere of the workshop and the wonderful 
hospitality in Cyprus.

\end{document}